\documentclass[aps,prd,twocolumn,superscriptaddress,amsmath,nofootinbib,natbib]{revtex4-1}

\usepackage[hidelinks]{hyperref}
\usepackage{amsfonts}
\usepackage{mathtools}
\usepackage{aas_macros}
\usepackage[capitalize]{cleveref}                                                             
\usepackage{lipsum}

\Crefname{equation}{Equation}{Equations}
\crefname{equation}{Eq.}{Eqs.}
\Crefname{figure}{Figure}{Figures}
\crefname{figure}{Fig.}{Figs.}
\Crefname{table}{Table}{Tables}
\crefname{table}{Tab.}{Tabs.}
\Crefname{section}{Section}{Sections}
\crefname{section}{Sec.}{Secs.}

\newcommand{\prm}[1]{{{#1}^{\prime}}}
\newcommand{\dprm}[1]{{#1}^{\prime\prime}}

\renewcommand{\d}[2][]{\operatorname{d}^{#1}\!{#2}}

\newcommand{\start}{\mathrm{i}}

\newcommand{\R}{\mathcal{R}}

\newcommand{\Planck}{\textit{Planck}}

\begin{document}

\title{Primordial power spectra for curved inflating universes}

\author{Will Handley}
\email[]{wh260@mrao.cam.ac.uk}
\affiliation{Astrophysics Group, Cavendish Laboratory, J.J.Thomson Avenue, Cambridge, CB3 0HE, UK}
\affiliation{Kavli Institute for Cosmology, Madingley Road, Cambridge, CB3 0HA, UK}
\affiliation{Gonville \& Caius College, Trinity Street, Cambridge, CB2 1TA, UK}
\homepage[]{https://www.kicc.cam.ac.uk/directory/wh260}

%

\date{\today}

\begin{abstract}
    Exact numerical primordial primordial power spectra are computed and plotted for the best-fit \Planck{} 2018 curved universe parameters. It is found that the spectra have generic cut-offs and oscillations within the observable window for the level of curvature allowed by current cosmic microwave background measurements and provide a better fit to current data. Derivations for the Mukhanov-Sasaki equation for curved universes are presented and analysed, and theoretical implications for the quantum and classical initial conditions for inflation are discussed within the curved regime. 
\end{abstract}

\maketitle

%
%

\section{Introduction}\label{sec:introduction}
Cosmological inflation~\cite{starobinskii_spectrum_1979,guth_inflationary_1981,linde_1982} is the current most popular theory for explaining the observed flatness and homogeneity of our present-day universe, while simultaneously providing a powerful framework for predicting the measured spectrum of anisotropies in the cosmic microwave background~\cite{planck_parameters,planck_inflation}. Nevertheless, small and unsatisfactory features in the cosmic microwave background (CMB) power spectra arguably still remain~\cite{planck_isotropy}, and there are ever-increasing tensions observed between datasets that measure the early universe and those that measure late-time properties~\cite{Riess2018, Joudaki2017, Kohlinger2017, Hildebrandt2017, DESParameters2017, tension}. The hunt is on to find extensions to the concordance cosmology (cosmological constant with cold dark matter --- $\Lambda$CDM) which are capable of resolving some or all of these discrepancies.

One extension that is often considered is to reintroduce a small amount of late-time curvature, creating a $K\Lambda$CDM cosmology~\cite{Ellis_2003,lasenbyclosed}. \Planck{} 2018 data without the lensing likelihood~\cite{planck_lensing} give relatively strong evidence for a closed universe~\cite{planck_parameters}. Adding in lensing and Baryon acoustic oscillation data~\cite{SDSS,SDSS2,SDSS3} reduces this evidence considerably, but it remains an open question as to why the CMB alone strongly prefers universes with positive spatial curvature (with possible implications for tension resolution).
Nevertheless, at the time of writing, universe models with percent-level spatial curvature remain compatible with current datasets.

There are theoretical reasons to consider the effect of curvature on the dynamics of inflation. If one is to invoke an inflationary phase in order to explain the observed present-day flatness, one cannot assume that the universe was flat at the start of inflation, and the presence of any curvature is arguably incompatible with eternal inflation. Furthermore, the observation of any amount of present-day curvature strongly constrains the total amount of inflation, providing a powerful justification for just-enough-inflation theories~\cite{kinetic_dominance,case_for_kd,Ramirez,justenough,BVS1,BVS2}

In the traditional curved $K\Lambda$CDM cosmology, a simple $(A_s, n_s)$ parametric form for the primordial power spectrum is usually assumed~\cite{planck_parameters,planck_inflation}:
\begin{equation}
    \mathcal{P}_\mathcal{R}^{K\Lambda\mathrm{CDM}}(k) = A_s \left( \frac{k}{k_*} \right)^{n_s-1}.
    \label{eqn:PPS_lcdm}
\end{equation}
In this work, I examine the effect on the fit of curved cosmologies to data if an exact numerical approach is used to calculate the primordial power spectrum. In all cases an improved fit is found.

In \cref{sec:mukhanov_sasaki} the Mukhanov-Sasaki equation is derived in the general case of curved universes and compared with the flat-space equivalent. In \cref{sec:actions} the general Mukhanov action is calculated and discussed with regards to its quantisation and consequent setting of initial conditions. In \cref{sec:pps} the primordial and CMB power spectra are calculated for the best-fit \Planck{} 2018 parameter values, and the fit is compared against the concordance case. Conclusions are presented in \cref{sec:conclusions}.

\section{The Mukhanov-Sasaki equation}\label{sec:mukhanov_sasaki}
In this section for completeness I derive the Mukhanov-Sasaki equation (\cref{eqn:mukhanov_sasaki}) for curved universes by a direct perturbative approach~\cite{Mukhanov_1992}. Similar computations have been performed historically by~\cite{De_Hai_2004,Gratton_2002,Ratra_2017,Bonga_2016,Bonga_2017,horndeski,Ooba_2017}. The analytical calculations throughout this paper were performed with the aid of computer algebra provided by Maple\textsuperscript{TM} 2017~\cite{maple,Maple10}, making use of the \texttt{Physics} and \texttt{DifferentialGeometry} packages. 

The action for a single-component scalar field minimally coupled to a curved spacetime is
\begin{equation}
    S = \int \d[4]{x}\sqrt{|g|}\left\{ \frac{1}{2}R + \frac{1}{2}\nabla^\mu\phi\nabla_\mu\phi - V(\phi)\right\}.
    \label{eqn:action}
\end{equation}
Extremising this action yields the Einstein field equations and a conserved stress energy tensor. Throughout this paper, in accordance with the cosmological principle, we shall assume that to zeroth order the solutions to these equations are homogeneous and isotropic. We then perturbatively expand these equations about the homogeneous solutions to first order in the Newtonian gauge, with the perturbation to the scalar field written as $\delta\phi$. In spherical polar coordinates the metric is therefore
\begin{gather}
    \d{s}^2 = (1+2\Phi)\d{t}^2 - {a(t)}^2 (1-2\Psi) (c_{ij}+h_{ij})\d{x}^i \d{x}^j,
    \nonumber\\
    c_{ij}\d{x^i}\d{x^j} = \frac{\d{r}^2}{1-Kr^2} + r^2(\d{\theta}^2  + \sin^2\theta\d{\phi}^2),
    \label{eqn:metric}
\end{gather}
where $K$ denotes the sign of the spatial curvature, taking values of $K=+1$ for a closed (positively curved) universe, $K=-1$ for an open (negatively curved) universe, and $K=0$ for a traditional flat universe. The covariant spatial derivative associated with the metric on comoving spatial slices is denoted with a Latin index as $\nabla_i$ with no factors of $a(t)$. The potentials $\Phi$ and $\Psi$ along with $\delta\phi$ are scalar perturbations whilst $h_{ij}$ is a divergenceless, traceless tensor perturbation with two independent polarisation degrees of freedom.

\subsection{Zeroth order equations}\label{sec:zero_order}
At zeroth order, the time-time component of the Einstein field equations and the time component of the conservation of the stress-energy tensor give the evolution equations of the homogeneous background fields
\begin{align}
    H^2 &=\frac{1}{3}\left(\frac{1}{2}\dot\phi^2 + V(\phi)\right)  - \frac{K}{a^2},
    \label{eqn:E00}\\
    0 &= \ddot\phi + 3 H \dot\phi + V'(\phi),
    \label{eqn:cons0}
\end{align}
where the Hubble parameter $H = \dot a / a$, and primes denote derivatives with respect to $\phi$. A further useful relation is
\begin{equation}
    \dot H = -\frac{1}{2}\dot\phi^2 + \frac{K}{a^2},
    \label{eqn:curv}
\end{equation}
which may be found be differentiating~\cref{eqn:E00} and eliminating the potential with~\cref{eqn:cons0}. \cref{eqn:E00,eqn:cons0} may be used to remove all explicit potential dependency from the first order equations and \cref{eqn:curv} can be used to remove all derivatives of $H$ in place of $\phi$, which is performed without comment in all of the below. 

\subsection{First order equations}\label{sec:first_order}
To first order, the time-time component of the Einstein field equations gives
\begin{equation}
    6 H \dot\Psi + 2 V \Phi + V' \delta \phi + \dot\phi \delta\dot\phi - \frac{2}{a^2}\nabla_i\nabla^i\Psi - 6 \frac{K}{a^2}(\Phi+\Psi) = 0,
    \label{eqn:dE00}
\end{equation}
the time-space components of the Einstein field equations all yield
\begin{equation}
    \dot\phi\: \delta \phi - 2 H \Phi - 2\dot\Psi = 0,
    \label{eqn:dEi0}
\end{equation}
the time component of the conservation equation shows
\begin{equation}                                      
    2V'\Phi  + V''\delta\phi -3\dot\phi\dot\Psi -\dot\phi\dot\Phi - \frac{1}{a^2}\nabla_i\nabla^i\delta\phi + \delta\ddot\phi + 3 H \delta\dot\phi = 0,
    \label{eqn:dcons0}
\end{equation}
the off-diagonal spatial components prove that
\begin{equation}
    \Phi=\Psi,
    \label{eqn:dEij}
\end{equation}
and the gauge-invariant comoving curvature perturbation is defined by the expression
\begin{equation}
    \mathcal{R} = \Psi +\frac{H}{\dot\phi}\delta\phi.
    \label{eqn:Req}
\end{equation}
Using the time derivative of \eqref{eqn:dEi0}, alongside \cref{eqn:dE00,eqn:dEi0,eqn:dcons0} we have four master equations. Substituting $\R$ for $\Phi$ and $\Psi$ into these using \cref{eqn:dEij,eqn:Req} allows us to eliminate $\delta\phi$ and its first and second time derivatives, yielding the Mukhanov-Sasaki equation
\begin{align}
    0=&(\mathcal{D}^2 - K\mathcal{E})\ddot{\R} + \left( \left( H + 2\frac{\dot z}{z} \right)\mathcal{D}^2 - 3K H \mathcal{E} \right)\dot{\R} \nonumber\\
    +&\frac{1}{a^2}\left( K\left(1+\mathcal{E} - \frac{2}{H}\frac{\dot{z}}{z}\right)\mathcal{D}^2 + K^2\mathcal{E}-\mathcal{D}^4 \right)\mathcal{R},
    \label{eqn:mukhanov_sasaki}
\end{align}
where
\begin{equation}
    \mathcal{D}^2 = \nabla_i\nabla^i + 3 K, \qquad
    z = \frac{a \dot{\phi}}{H}, \qquad
    \mathcal{E} = \frac{\dot{\phi}^2}{2H^2}.
    \label{eqn:definitions}
\end{equation}
Upon Fourier decomposition, one simply replaces the $\mathcal{D}^2$ operator in \cref{eqn:mukhanov_sasaki} with its associated scalar wavevector expression~\cite{fouriercurved}
\begin{align}
    \mathcal{D}^2 &\leftrightarrow -k^2+3K,    & k \in \mathbb{R}, k > 0:& & K&=0,-1, \nonumber\\
    \mathcal{D}^2 &\leftrightarrow -k(k+2)+3K, & k\in\mathbb{Z}, k > 2:&   & K&=+1,\label{eqn:wavevectors}
\end{align}
One may interpret the operator $\mathcal{D}^2$ physically by examining the Ricci three-scalar to first order
\begin{equation}
    R^{(3)} = 6\frac{K}{a^2} + \frac{4}{a^2}(c^{ij}\nabla_i\nabla_j + 3K)\R,
\end{equation}
so the perturbation to comoving spatial curvature can be seen by inspection to be $\frac{4}{a^2}\mathcal{D}^2\R$.

The Mukhanov-Sasaki~\cref{eqn:mukhanov_sasaki} in the curved case demands some comment. First, it should be noted that for the flat case $K=0$ it collapses down to its usual form, albeit with an additional $\nabla_i\nabla^i$ multiplying each term. The same observations apply for the small-scale limit $k\to\infty$. The addition of curvature considerably increases the complexity of the evolution equations at low and intermediate $k$ by adding wavevector-dependent coefficients to all three terms in front of $\ddot{\R}$, $\dot{\R}$ and $\R$. As we shall see in \cref{sec:pps}, this has consequences for the evolution of the comoving curvature perturbation and the resulting primordial power spectrum.

The tensor equivalent to \cref{eqn:mukhanov_sasaki} is derived similarly, yielding
\begin{equation}
    \ddot{h}  + 3 H \dot{h} - \frac{1}{a^2}\left( \nabla_i\nabla^i - 2K \right)h = 0,
    \label{eqn:mukhanov_sasaki_tensor}
\end{equation}
for both polarisation modes of the tensor perturbation.
Here the modification provided by curvature is significantly simpler, and readers can confirm that it reduces to the flat-space equivalent in the case that $K=0$ and for $k\gg1$.

\section{The Mukhanov action}\label{sec:actions}
In this section I confirm the calculation in \cref{sec:mukhanov_sasaki} by arriving at \cref{eqn:mukhanov_sasaki} via the Mukhanov action. I follow the notation of \citet[Appendix B]{tasi}, generalising their calculation to the curved case. 

The simplest approach for deriving the perturbation to the action in \cref{eqn:action} is to write the metric in the ADM formalism~\cite{adm,gauge_invariant}, where spacetime is sliced into three dimensional hypersurfaces
\begin{equation}
    \d{s}^2 =  - N^2 \d{t}^2 + g_{ij}^{(3)} (\d{x^i} + N^i\d{t})(\d{x^j} + N^j\d{t}).
\end{equation}
With the metric in this form, we find that the action from \cref{eqn:action} becomes
\begin{align}     
    S = \frac{1}{2}&\int\d[4]{x}\sqrt{|g^{(3)}|}
    [ NR^{(3)} + N^{-1}(E_{ij}E^{ij} - E^2) \nonumber\\
    &+ N^{-1}{(\dot{\phi}-N^i\nabla_i\phi)}^2 - N \nabla_i\phi\nabla^i\phi- 2NV],\label{eqn:ADM}\\
    E_{ij} &= \frac{1}{2}(\dot{g}_{ij}^{(3)} - \nabla_i N_j - \nabla_j N_i),\qquad E = E^i_i.
\end{align}
Focussing on first order scalar perturbations, we write
\begin{equation}
    N = 1+\alpha,\qquad N_i = \nabla_i\psi.
\end{equation}
Working in the comoving gauge where $\delta\phi=0$,the spatial part of the ADM metric is defined to be
\begin{equation}
    g_{ij}^{(3)} = a^2 (1-2\R)c_{ij}.
\end{equation}
The Lagrangian constraint equations are
\begin{align}
    \alpha &= -\frac{\dot{\R}}{H} + \frac{K}{a^2}\frac{\psi}{H},
    \label{eqn:alpha}\\
    \frac{1}{a^2H}\mathcal{D}^2\R-\mathcal{E}\dot{\R} &= \frac{1}{a^2}\mathcal{D}^2 \psi - \frac{K}{a^2}\mathcal{E}\psi.
    \label{eqn:psi}
\end{align}
We may formally solve \cref{eqn:psi} explicitly for $\psi$ with the rather cryptic expression
\begin{equation}                                                            
    \psi =\frac{\R}{H} -a^2\mathcal{E}{(\mathcal{D}^2 -K\mathcal{E})}^{-1}(\dot{\R}  - \frac{K}{a^2} \frac{\R}{H}).
    \label{eqn:psi_eq}
\end{equation}
By the construction of the ADM formalism, substituting the first order solutions from \cref{eqn:alpha,eqn:psi} into the action from \cref{eqn:ADM} gives the second order action. After some effort integrating this by parts, we find
\begin{align}
    S = \frac{1}{2}\int \d[4]{x}\sqrt{|c|}a^3 &\frac{\dot{\phi}^2}{H^2}\Bigg\{{\left( \dot\R - \frac{K}{a^2H}\R \right)}^2 \nonumber\\
     & - \frac{K}{a^2}\left( \dot\R - \frac{K}{a^2H}\R \right)\left(\psi-\frac{\R}{H}\right)
     \nonumber\\
     &- \frac{1}{a^2}\nabla_i\R\nabla^i\R  + 3\frac{K}{a^2}\R^2
 \Bigg\}.
\end{align}
Substituting \cref{eqn:psi_eq} into the above, and integrating by parts one more time returns the unusual action
\begin{align}
    \frac{1}{2}\int& \d[4]{x}\sqrt{|c|}a^3 \frac{\dot{\phi}^2}{H^2}\Bigg\{\frac{1}{a^2}\R\mathcal{D}^2\R
     \nonumber\\
 &+ \left(\dot{\R}  - \frac{K}{a^2} \frac{\R}{H}\right)\frac{\mathcal{D}^2}{\mathcal{D}^2 -K\mathcal{E}}\left(\dot{\R} -\frac{K}{a^2}\frac{\R}{H}\right)\Bigg\}.
\end{align}
Varying this action with respect to $\mathcal{R}$ recovers the Mukhanov-Sasaki equation from \cref{eqn:mukhanov_sasaki}.

The full curved action is worthy of comment. Setting $K=0$ recovers the flat-space action, but with non-zero $K$ the action becomes non-local due to the presence of a denominator with a derivative term. 

In the flat case, the usual next step is to diagonalise the action so that it has a canonical normalisation\footnote{A canonically normalised quantum field $\phi$ with mass $m$ has action of the form {$S = \int\d{t}\d[3]{x}\:\dot\phi^2 + \phi(-\nabla^2 + m^2)\phi$}.} by transforming to conformal time $\d{\eta} = a \d{t}$ and re-phrasing in terms of the Mukhanov variable $v=z\R$. In the curved case, this is impossible. The best one can do is to define a wavevector-dependent $\mathcal{Z}$ and $v$ via:
\begin{equation}
    v = \mathcal{Z}\R, \qquad \mathcal{Z} = \frac{a\dot{\phi}}{H} \sqrt{\frac{\mathcal{D}^2}{\mathcal{D}^2 -K\mathcal{E}}},
\end{equation}
in which case, the action becomes
\begin{align}            
    \frac{1}{2}\int \d{\eta}\d[3]{x}\sqrt{|c|}\Big(&\prm{v}^2 - {(\nabla v)}^2 
        \nonumber\\
        &+ \left(\frac{\dprm{\mathcal{Z}}}{\mathcal{Z}} + 2K+\frac{2K\prm{\mathcal{Z}}}{\mathcal{H}\mathcal{Z}} \right)v^2\Big).
\end{align}
The lack of canonical normalisation implied by the $k$-dependent $\mathcal{Z}$ has theoretical consequences for the initial conditions for inflation, since for low to intermediate $k$ one cannot draw an analogy to the de-Sitter case in order to define initial conditions. The correct theoretical choice for initial conditions in this case is far from clear, and it may be that the only way to differentiate between competing approaches is to choose the correct initial conditions via confrontation with data.

The tensor part of the action is
\begin{equation}
    \frac{1}{16}\int \d[4]{x}\sqrt{|c|}a^3 \left(
        \dot{h}_{ij} \dot{h}^{ij} + \frac{1}{a^2}h_{ij}(\nabla_k\nabla^k -2K) h_{ij}
    \right),
\end{equation}
which even in the presence of curvature remains canonically quantisable in the traditional sense by switching to conformal time and making the variable redefinition $v_\mathrm{t} =ah$.

\begin{figure*}                                                                           
    \includegraphics{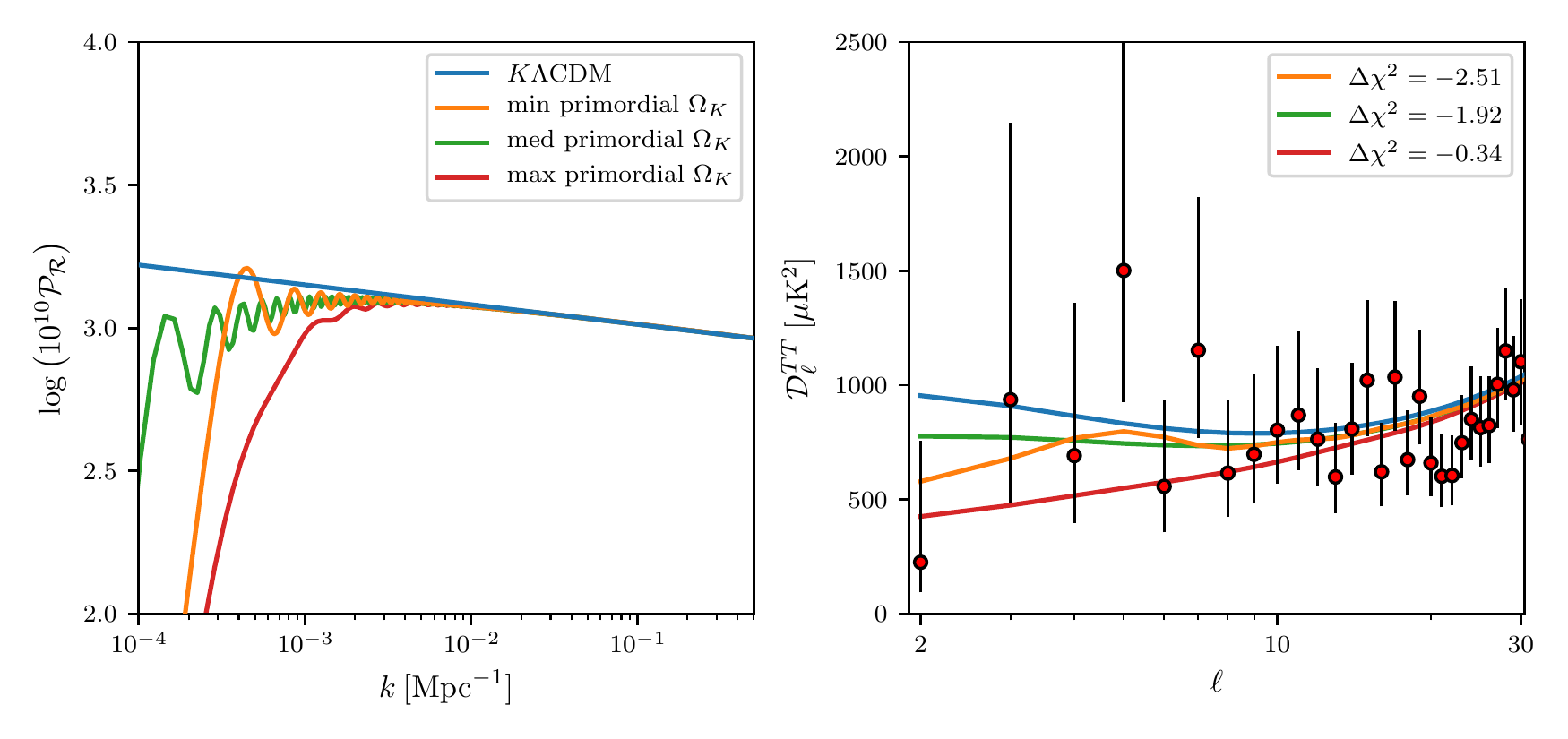}
    \caption{Left: representative best-fit primordial power spectra corresponding to the range of allowed primordial curvatures. Oscillations and a generic suppression of power are visible at low-$k$. The jagged edges of the curves at low-$k$ arise from the discreteness  of the wavevectors for closed universes indicated in \cref{eqn:wavevectors}. Right: the corresponding low-$\ell$ effects on the CMB power spectrum. The improvement in $\Delta\chi^2$ relative to $K\Lambda$CDM is shown in the right-hand figure legend, with negative values indicating a better fit to the data. The best-fit spectra without including the full primordial power spectrum calculation is highlighted in blue. Plots are shown for the entire observational window for Planck-like data on the left, and the plot on the right highlights the deviating region for the CMB power spectrum. There is no appreciable deviation from the traditional power spectrum at higher $k$ and $\ell$ values.\label{fig:BD}}
\end{figure*}
\begin{figure*}
    \includegraphics{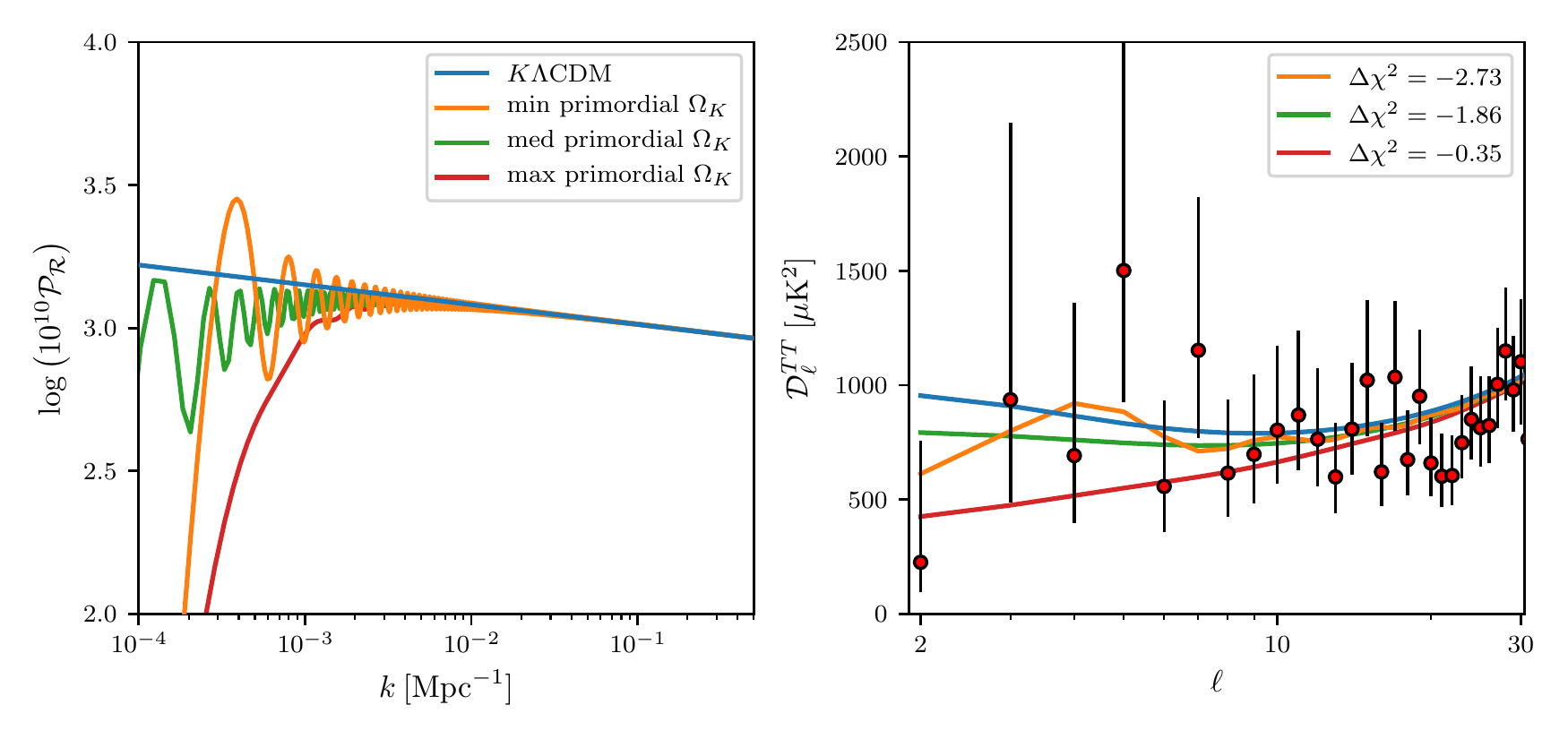}
    \caption{Same as \cref{fig:BD}, but using RST initial conditions instead of BD from \cref{eqn:IC}. The oscillations in the primordial power spectrum (left panel) are enhanced by RST initial conditions, resulting in a change in the $\Delta\chi^2$ for all cases, and a marginally better fit for the best-fitting case (right panel).\label{fig:RST} 
    }
\end{figure*}
\begin{figure*}
    \includegraphics{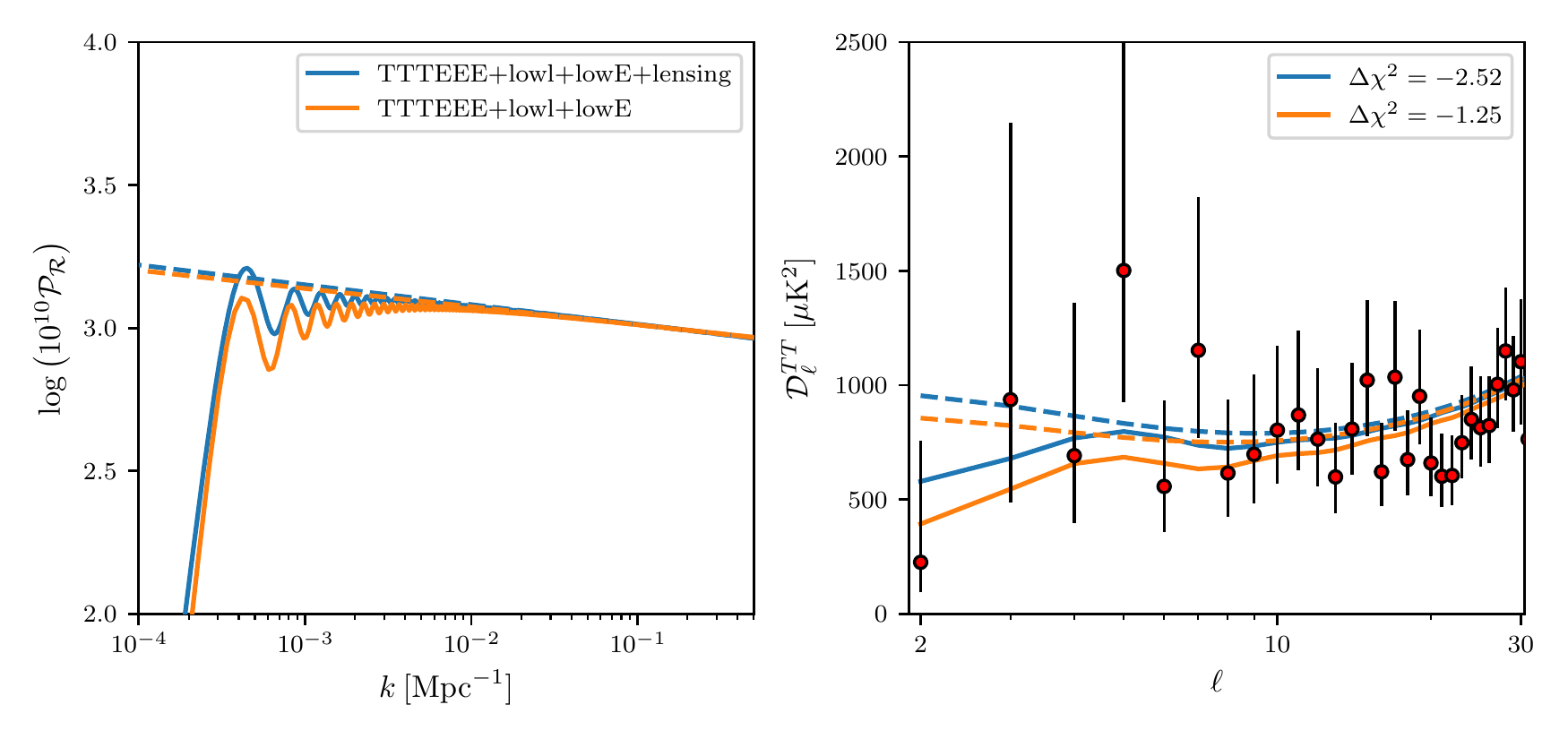}
    \caption{Primordial power spectra varying the amount of late time curvature. With CMB lensing data, the best-fit cosmology has $\Omega_K=-0.9\%$, whilst without CMB lensing $\Omega_K=-4.5\%$. Varying the amount of late time to this degree does not affect the location of the cutoff, but does adjust the suppression of power. The inclusion of the full numerical power spectrum in both cases results in an improved fit. Initial conditions are Bunch-Davies and both are chosen to have the ``minimum primordial curvature''.\label{fig:BD_nolenscomp} 
    }
\end{figure*}
\pagebreak
\section{The primordial power spectrum}\label{sec:pps}
To compute the exact primordial power spectrum, the Mukhanov variable $\R$ is evolved via the Mukhanov-Sasaki equation \cref{eqn:mukhanov_sasaki} for a relevant range of wavevectors $k$. The primordial power spectrum is then given by the limiting value of $\R$ after horizon exit
\begin{equation}
    \mathcal{P}_\mathcal{R}(k) = \lim_{k\ll aH}\frac{k^3}{2\pi^2} |\mathcal{R}|^2.
    \label{eqn:PR}
\end{equation}
The initial conditions for the evolution of the Mukhanov variable represent the injection of quantum mechanics into this system. For simplicity, I consider two types of initial conditions: Bunch--Davies (BD) and Renormalised stress-energy tensor (RST), which are set at inflation start $t_\start$:
\begin{equation}
    \R_\start = \frac{1}{\sqrt{2k}z_\start},\qquad
    \frac{\dot\R_\start}{\R_\start} = \left\{
    \begin{array}{cl}
     -i\frac{k}{a_\start} - \frac{\dot{z_\start}}{z_\start} & \mathrm{BD}\\
     -i\frac{k}{a_\start}                     & \mathrm{RST}\\
    \end{array}
    \right.
    \label{eqn:IC}
\end{equation}
BD initial conditions are theoretically only appropriate for a de-Sitter like spacetime, which at inflation start for low-$k$ modes is not true, whilst RST initial conditions are designed to be valid in all regimes~\cite{RST}.

The numerical integration itself is most efficiently performed using a solver that is capable of accurately navigating the many oscillations between initial conditions and horizon exit. Such solvers have undergone recent development~\cite{RKWKB,Haddadin,Agocs,Bamber} and in this work I use the latest of these provided by \citet{Agocs}.

For the evolution of the background variables I assume a \Planck{} 2018 TTTEEE+lowl+lowE+lensing best-fit concordance $K\Lambda$CDM curved cosmology, a monomial inflaton potential $V(\phi) \propto \phi^{4/3}$ and a reheating phase modelled by continuing the inflaton evolution until intersection with late-time Friedmann horizon. Interestingly, under this background setup, chaotic and Starobinsky potentials are incompatible with the \Planck{} best-fit cosmology when curvature is included.
The full pipeline of how these background evolutions are constructed will be discussed in an upcoming paper~\cite{curvedconstraints}. The $K\Lambda$CDM parameters pin down all but one degree of freedom in the background evolution, leaving a single primordial parameter determining the degree of primordial curvature at the start of inflation, or equivalently the scale factor of the universe at inflation start $a_\mathrm{i}$. For the curved case, the size of the universe at the start of inflation is bounded from below by the requirement that $H_\mathrm{i}>0$ and bounded from above by the requirement that the horizon problem is solved, i.e.\ that the amount of conformal time before inflation is greater than the amount of time afterwards. This amounts to a constraint that $22\mathrm{k}\ell_\mathrm{p}<a_\mathrm{i}<171\mathrm{k}\ell_\mathrm{p}$. Beyond the lower bound the primordial curvature diverges as $H\to 0$ and the universe begins in an emergent coasting state~\cite{Labra_2016,Ellis_2004}. Solutions for this case however are incompatible with the Planck best-fit parameters. The minimum amount of primordial curvature at the upper bound is $-2.1\%$, the maximum amount of curvature is $\sim-150$, and the ``medium'' amount of curvature in the figures is the geometric mean of these two.

The primordial power spectrum for the BD case is plotted in \cref{fig:BD}, and for RST initial conditions in \cref{fig:RST}. Relative to the concordance $K\Lambda$CDM, which assumes the almost flat power spectrum from \cref{eqn:PPS_lcdm}, including the exact numerical calculation introduces oscillations and a suppression of power at low $k$, independent of initial conditions. Varying the remaining degree freedom provided by the amount of primordial curvature alters the oscillations and level of suppression in a non-monotonic manner.

In both \cref{fig:BD,fig:RST} these predictions for the primordial power spectrum are followed through to the CMB~\cite{class}. For all allowed values of initial primordial curvature, incorporating the exact numerical solution results in an improved $\Delta\chi^2$ relative to $K\Lambda$CDM. Furthermore, the data are capable of distinguishing a preferred vacuum state, with the best fit preferring RST initial conditions over the traditional Bunch--Davies vacuum.

It should be noted that these $\Delta\chi^2$ values are not derived from a true fitting procedure. First, in the absence of a publicly available likelihood at the time of writing, the approximate $\Delta\chi^2$ value is computed from the available compressed $C_\ell$ spectra and their error bars. Second (for similar reasons) I have used the best-fit cosmological parameters derived from the $K\Lambda$CDM data with the default spectrum, rather than a full fit with the modified power spectrum. Given the degeneracies between cosmological parameters it is possible that the $\Delta\chi^2$ could be significantly enhanced under a full fitting. 

This work will be followed by paper detailing a Bayesian fit~\cite{constraining_kd} for these $K\Lambda$CDM universes with exact power spectra~\cite{curvedconstraints}. It remains to be seen whether the improved $\Delta\chi^2$ will be strong enough to balance the Occam penalty arising from the introduction of an additional constrained primordial parameter.

To see how these results vary as the amount of late-time curvature is altered in \cref{fig:BD_nolenscomp} the fit from \cref{fig:BD} is compared with the corresponding fit for the Planck data excluding CMB lensing, which has a significantly higher curvature of $\Omega_K=-4.5\%$ (as opposed to $-0.9\%$). In this case the location of the suppression of power and the oscillations are barely changed, although as expected~\cite{lasenbyclosed} the depth of the suppression is greater for the case with more negative curvature.

It should be noted that at the time of writing there is renewed interest in curved cosmologies in light of the controversial work by \citet{valentino} (and also \cite{curvature_tension}). In these analyses it is pointed out the Planck CMB primary data are in 2.5$\sigma$ and 3$\sigma$ tension with CMB lensing and baryon acoustic oscillation (BAO) data if curvature is included as a free parameter. Such results do not prove that the universe is curved, but arguably weaken the evidence for a flat universe since inconsistent datasets should not be combined. The cause of this tension could be an improbable statistical fluctuation, or could indicate a systematic error in one or more of the datasets. For example, it has been suggested recently that whilst BAO likelihoods are curvature agnostic, the compression strategies applied to the observational data do depend on curvature~\cite{BAO}, and the default is to assume a flat cosmology for performing this compression. These kind of baked-in flatness assumptions (if present) could bias the curvature constraint provided by BAO. Similarly, the CMB lensing likelihood expands to first order about a fiducial flat cosmology, and a full cross-check expanding about the best-fit $\Omega_K=-4.5\%$ cosmology has not yet been performed.

\section{Conclusions}\label{sec:conclusions}

In this work, the Mukhanov-Sasaki equations and actions for curved cosmologies were derived and discussed. 
It was found that including an exact numerical calculation for the primordial power spectrum gives a better fit to the data, and that current datasets are capable of distinguishing between alternative definitions of the quantum vacuum. It remains to be seen whether a more complete Bayesian fitting procedure yields compelling evidence for universes with curvature, or the ability to distinguish quantum vacua.

\begin{acknowledgements}
    I am indebted to Anthony Lasenby, Mike Hobson, Lukas Hergt, Fruzsina Agocs, Denis Werth and Maxime Jabarian for their many conversations and participation in the ongoing theoretical and observational research programming stemming from this work.
I thank Gonville \& Caius College for their continuing support via a Research Fellowship. 
\end{acknowledgements}
\vfill

\bibliographystyle{unsrtnat}
\bibliography{mukhanov_sasaki}
\end{document}